\documentclass{article}

\begin{document}

\begin{center}

{\large \textbf{FIELD THEORIES ON CANONICAL AND LIE-ALGEBRA\\
\vspace{0.3cm} NONCOMMUTATIVE SPACETIMES}\footnote{Invited talk
given by G.~Amelino-Camelia at the {\it 25th Johns Hopkins Workshop on Current
Problems in Particle Theory}, Florence,  September 3-5, 2001 }}

\vspace{1.1cm}

{\bf Giovanni AMELINO-CAMELIA, Michele ARZANO and Luisa DOPLICHER}

\vspace{0.6cm}

\textit{Dipart.~Fisica, Univ.~Roma ``La Sapienza'', P.le Moro 2,
\\00185 Roma, Italy\\E-mail: amelino@roma1.infn.it,
arzano@roma1.infn.it, doplicher@roma1.infn.it}

\end{center}

\begin{abstract}
Field theories on canonical noncommutative spacetimes, which are
being studied also in connection with string theory, and on
$\kappa$-Minkowski spacetime, which is a popular example of
Lie-algebra noncommutative spacetime, can be naturally
constructed by introducing a suitable generating functional for
Green functions in energy-momentum space. Direct reference to a
star product is not necessary. It is sufficient to make use of
the simple properties that the Fourier transform preserves in
these spacetimes and establish the rules for products of wave
exponentials that are dictated by the non-commutativity of the
coordinates. The approach also provides an elementary description
of ``planar" and ``non-planar" Feynman diagrams. We also comment
on the rich phenomenology emerging from the analysis of these
theories.
\end{abstract}

Noncommutative geometry is being used more and more extensively in attempts
to unify general relativity and quantum mechanics. Some ``quantum-gravity"
approaches explore the possibility that noncommutative geometry might
provide the correct fundamental description of spacetime, while in other
approaches noncommutative geometry turns out to play a role at the level of
the effective theories that describe certain aspects of quantum gravity.

Two simple examples~\cite{starwess} are ``canonical noncommutative
spacetimes" ($\mu,\nu,\beta = 0,1,2,3$)
\begin{equation}
\left[x_\mu,x_\nu\right] = i \theta_{\mu \nu}
\label{canodef}
\end{equation}
and ``Lie-algebra noncommutative spacetimes"
\begin{equation}
\left[x_\mu,x_\nu\right] = i C^\beta_{\mu \nu} x_\beta ~.
\label{liedef}
\end{equation}
The canonical type (\ref{canodef}) was originally proposed~\cite{dopl} in
the context of attempts to develop a new fundamental picture of spacetime.
More recently, (\ref{canodef}) is proving useful in the description of
string theory in certain $B$-field backgrounds (see, {\it e.g.},
Refs.~\cite{seibwitt,sussIRUV,chong}), with the tensor $\theta_{\mu \nu}$
reflecting the properties of the background.

Among the Lie-algebra spacetimes (\ref{liedef}), one of the most studied as
an alternative to classical Minkowski spacetime is the
$\kappa$-Minkowski~\cite{majrue,kpoinap} spacetime ($l,m = 1,2,3$)
\begin{equation}
\left[x_m,t\right] = {i \over \kappa} x_m ~,~~~~\left[x_m, x_l\right] = 0 ~.
\label{kmindef}
\end{equation}
One of us recently proposed~\cite{dsr1dsr2} a new path toward quantum
gravity in which $\kappa$-Minkowski would play a key role.

In this paper we aim at giving a short introduction to the
construction of field theories in canonical and
$\kappa$-Minkowski (Lie-algebra) noncommutative spacetimes,
showing that the same strategy can be adopted in both cases.

Building a field theory on these spacetimes will require writing down a
generating functional for Green functions that involves products of
noncommuting fields. The first observation concerns the way to handle
functions of the (noncommuting) spacetime variables. A key point is that
one can define such functions as inverse Fourier transforms of some
ordinary/commutative energy-momentum-space functions.\footnote{The reader
will recognize this as a Weyl quantization,~\cite{starwess,castellani}
which was originally introduced for the noncommuting phase space variables
of ordinary quantum mechanics.} Our functions in noncommutative spacetime
will be written as\footnote{Note that, since we always refer to
noncommutative spacetime coordinates unless explicitly clarified in the
text, in our equations we do not adopt a special notation (such as
$\hat{x}$, which is adopted by some authors) for noncommuting spacetime
coordinates. Also note that throughout the paper energy-momentum-space
coordinates are always commutative.}
\begin{equation}
f\left( x\right)
=\frac{1}{\left( 2\pi \right) ^{2}}\int d^{4}k
~:\exp \left(ik^{\mu } x_{\mu }\right): ~\tilde{f}\left( k\right)\quad,
\label{FT}
\end{equation}
where $\tilde{f}\left( k\right)$ is the Fourier transform of $f$ written in
terms of the commuting coordinates and the Fourier parameters $k_{\mu }$
are commuting variables.
The function $:\exp \left(ik^{\mu } x_{\mu}\right):$
must be consistent with the properties~\cite{majidfourier} of
the Fourier transform and must reproduce the
ordinary $\exp \left(ik^{\mu } x_{\mu }\right)$
in the commutative limit.
In canonical spacetimes one can take~\cite{starwess}
for $:\exp \left(ik^{\mu } x_{\mu}\right):$
simply $\exp \left(ik^{\mu } x_{\mu }\right)$ (here of course
intended as a function of noncommuting spacetime coordinates and commuting
Fourier parameters),
\begin{equation}
:e^{i k^\mu x_\mu}:_\theta \equiv e^{i k^{\mu } x_{\mu }}
~,
\label{canorder}
\end{equation}
whereas it turns out~\cite{gacmaj,lukistar,majidfourier} that in the
$\kappa$-Minkowski case the correct formulation of the Fourier theory
requires a suitable ordering prescription:\footnote{There is of course an
equally valid alternative ordering prescription in which the time-dependent
exponential is placed to the left~\cite{lukistar} (while we are here
choosing the convention with the time-dependent exponential to the right).}
\begin{equation}
:e^{i k^\mu x_\mu}:_\kappa \equiv e^{i k^m x_m} e^{i k^0 x_0}
~.
\label{order}
\end{equation}
When,
in Eq.~(\ref{FT}),
both $f(x)$ and the phase exponential are ordered as in (\ref{order}),
the function $\tilde{f}\left( k\right)$ is again the Fourier
transform of $f$ written in terms of the commuting coordinates.

By introducing the Fourier transform one reduces the problem
of describing products of
fields to the one of establishing the
properties of products of exponentials of the noncommuting variables.
These can often be evaluated straightforwardly ({\it e.g.} using
the Baker-Campbell-Hausdorff formula). In the
canonical case one finds immediately:
\begin{equation}
\left ( e^{i p^\mu x_\mu} e^{i k^\nu x_\nu} \right )_\theta
= e^{-\frac{i}{2} p^\mu
\theta_{\mu \nu} k^\nu} e^{i (p+k)^\mu x_\mu} ~,
\label{expprodcano}
\end{equation}
{\it i.e.} the Fourier parameters $p_\mu$ and $k_\mu$ combine just as
usual, with the only new ingredient of an overall phase factor that depends
on $\theta_{\mu \nu}$. On the $\kappa$-Minkowski side one introduces
ordered exponentials primarily as a way to obtain simple
rules for the product of wave exponentials: while wave exponentials of the
type $e^{i p^\mu x_\mu}$ would not combine in a simple way,
for the ordered wave exponentials one finds
\begin{equation}
(:e^{i p^\mu x_\mu}:_\kappa) (:e^{i k^\nu x_\nu}:_\kappa) =
:e^{i (p \dot{+} k)^\mu x_\mu}:_\kappa
\quad.
\label{expprodlie}
\end{equation}
The notation ``$\dot{+}$" here introduced reflects the
behaviour of the composition of momenta in $\kappa$-Minkowski
spacetime. The deformed addition is not symmetric in the two
momenta; it works as follows (no sum on repeated indices):
\begin{equation}
p_\mu \dot{+} k_\mu \equiv \delta_{\mu,0}(p_0+k_0) + (1-\delta_{\mu,0})
(p_\mu +e^{p_0/\kappa} k_\mu) ~. \label{coprod}
\end{equation}
The element $\dot{-} p$ such that $p \dot{+} (\dot{-} p) =0$
is accordingly defined as follows:
\begin{equation}
(\dot{-} p)_\mu \equiv \delta_{\mu,0}(-p_0) + (1-\delta_{\mu,0})
(-e^{-p_0/\kappa} p_\mu) ~.
\label{inverse}
\end{equation}
Readers familiar with the $\kappa$-Poincar\'{e} research programme will
recognize in (\ref{coprod}) the rule for the energy-momentum ``coproduct"
and in (\ref{inverse}) the rule for the ``antipode". From (\ref{coprod})
one can see that the energy\footnote{We are here taking the liberty to
denominate ``energy" the Fourier parameter in the 0-th direction (and
similarly for the other three Fourier parameters we use ``3-momentum").
This terminology may appear unjustified in the present context, but it is
actually meaningful in light of the results on $\kappa$-Minkowski reported
in Ref.~\cite{gacmaj}.} addition is undeformed while 3-momenta are added
according to $\vec{p}+e^{p_0/\kappa}\vec{k}$.

In our approach to the construction of these field theories the
noncommutativity of the spacetime coordinates is only reflected in the
properties of products of wave exponentials, (\ref{expprodcano}) and
(\ref{expprodlie}). In similar studies of field theory in noncommutative
spacetime one often takes one additional step, by introducing the so called
``star-product": the algebra of functions of noncommuting variables is
represented by an algebra of ordinary functions endowed with a deformed
product. However, when aiming to a formulation of the field theory (Feynman
rules) in energy-momentum space, one only needs to handle products of
exponentials of the coordinates, and there is no need to make explicit
reference to a star product (which would allow one to handle general
products of functions of the coordinates). The observations
(\ref{expprodcano}) and (\ref{expprodlie}) on products of wave exponentials
are all one needs (they of course are at the root of the star product, as
one can easily realize by appropriate use of Fourier transforms).

 From (\ref{expprodcano}) and (\ref{expprodlie}) it is possible to
deduce some important features of field theories on a
noncommutative spacetime. The noncommutativity of the coordinates
of course makes the product of fields noncommutative, and this
could be alarming with respect to some of the most significant
symmetry properties of Feynman diagrams (such as symmetries under
external-line exchange when identical particles are involved).
Particularly for $\kappa$-Minkowski it was
suspected~\cite{gacmaj,lukistar,lukiFT} that this difficulty,
reflected in the lack of symmetry of the coproduct
(\ref{coprod}), would lead to nonsensical results. However, it
turns out~\cite{gacmichele} that this expectation is incorrect.
In the following we will show in detail, in the case of a scalar
theory with quartic interaction (``$\lambda \Phi^4$ theory"),
that both in the canonical case and in the $\kappa$-Minkowski
case the new features introduced by the noncommutativity do not
spoil the desired symmetries under exchange of identical
particles.

We start with a generating functional for Green
functions (partition function):
\begin{eqnarray}
Z\left[J \right]&=&\int {\mathcal D}\left[\phi\right]\,\exp \left( i \int
 d^4x \, \left[\frac{1}{2} \partial^{\mu}\phi(x)\partial_{\mu}\phi(x)
-\frac{m^2}{2} \phi^2(x) \right. \right. \nonumber \\
&&\left. \left. - \frac{\lambda}{24}\phi^4(x) +\frac{1}{2} J(x)\phi(x)+
\frac{1}{2} \phi(x)J(x)\right] \right)~.
\label{zeta1st}
\end{eqnarray}
This expression can be maintained in both our examples
of noncommutative spacetime.
We intend to obtain energy-momentum-space Feynman rules,
so we must rewrite the partition function in energy-momentum space,
using the representation (\ref{FT}).
We also need an
integral representation of the delta function
and some elements of the differential calculus
in the noncommutative spacetime.

We denote the
delta function by $\delta^{(4)}_{D}(k)$, where the subscript $D$ is
replaced by $\theta$ on the canonical side and by $\kappa$ on the
$\kappa$-Minkowski side. In the canonical case we write:
\begin{equation}
\delta^{(4)}_{\theta}(k)=\int \frac{d^4x}{(2\pi)^4}\exp(ikx) ,
\end{equation}
in very close analogy\footnote{The careful reader should notice that
throughout our analysis formulas for the canonical spacetime are closer to
the ones for the familiar case of commutative spacetime. The
$\kappa$-Minkowski side requires instead more care. On the canonical side
certain simplifications are possible because of the fact that commutators
of coordinates are coordinate independent.} with the familiar commutative
spacetime delta function, while in $\kappa$-Minkowski spacetime we need
again an ordered exponential:~\cite{lukistar}
\begin{equation}
\delta^{(4)}_{\kappa}(k)=\int\frac{d^4x}{(2\pi)^4}:\exp(ikx): \quad.
\end{equation}
The manipulations of the partition function on the $\kappa$-Minkowski side
will lead to the emergence of terms like $\delta^{(4)}(p \dot{+} k)$. The
lack of symmetry of $p \dot{+} k$ then
requires some care; the relevant formulas are
\begin{eqnarray}
\int d^4k \, \delta^{(4)}_{D}(k\dot{+}p)f(k)&=& \chi_1(-p_0) \,
f(\dot{-}p) ~, \label{deltaone} \\
\int d^4k \, \delta^{(4)}_{D}(p\dot{+}k)f(k)&=& \chi_2(p_0) \,
f(\dot{-}p) ~, \label{deltatwo}
\end{eqnarray}
where $\chi_1(p_0)=\chi_2(p_0)=1$ in the canonical case and
\begin{eqnarray}
\chi_1(p_0)&=& \mu(p_0) \\
\chi_2(p_0)&=&e^{-3 p_0/\kappa} \, \mu(p_0) \label{chi}
\end{eqnarray}
for $\kappa$-Minkowski.
The function $\mu(p_0)$, which
is not needed on the canonical side, is intended~\cite{gacmichele} to
reflect the properties of a non-trivial measure of integration over
$\kappa$-energy-momentum space (which, in the sense reflected by
(\ref{coprod}), is not flat). On $\mu(p_0)$ we will only observe and use
the fact that it can depend on energy-momentum only through the 0-th
component (energy),~\cite{gacmaj} postponing to future studies the
determination of its exact form.
Concerns about the conservation rules have been the most serious obstacle
for the construction of physical theories based on $\kappa$-Minkowski, and
we shall show that these concerns can be straightforwardly addressed within
our approach, independently of the form of $\mu(p_0)$.

On the $\kappa$-Minkowski side one also needs~\cite{gacmaj} a suitable
nontrivial definition of partial derivative:
\begin{eqnarray}
{\partial \over \partial x_m} :e^{i p  x }: &= &: {\partial \over \partial
x_m} e^{i p  x }: \label{partialderiv1} \\ {\partial \over \partial x_0}
:e^{i p  x }: &=&
\kappa : \left( e^{i p  x }
- e^{i p  (x + \Delta x_\kappa)} \right): ~,
\label{partialderiv2}
\end{eqnarray}
where $(\Delta x_\kappa)_\mu \equiv - \delta_{\mu,0}/\kappa$.

Our first objective is to examine the structure of the two-point function
at tree-level. For this result we can of course switch off the coupling
$\lambda$. Using (\ref{deltaone}), (\ref{deltatwo}), (\ref{partialderiv1})
and (\ref{partialderiv2}) one obtains from (\ref{zeta1st}) the (momentum
space) deformed partition function of the free theory:
\begin{eqnarray}
Z^0_D\left[J \right] &=&\int {\mathcal D}\left[\phi\right] \exp
\left(\frac{i}{2}\int d^4k \chi_1(k_0) \left[\phi(\dot{-}k)
\left[{\mathcal C}_{D}(k)-m^2\right] \phi(k)\right. \right.
\nonumber \\ &&\left. \left.
+J(k)\phi(\dot{-}k)+\phi(k)J(\dot{-}k)\right]\right)
\label{zeta2nd}\quad,
\end{eqnarray}
where in the canonical case the energy-momentum
coproduct is trivial, $(\dot{+})_\theta \equiv +$,
and accordingly also the energy-momentum antipode
is trivial, $(\dot{-})_\theta \equiv -$.
The form of $\chi_1(k_0)$ depends on the kind of noncommutativity (as
already specified), and ${\mathcal C}_{D}$ represents the mass casimir,
{\it i.e.} in the canonical case
\begin{equation}
{\mathcal C}_{\theta}(k) = k_0^2-\vec{k}^2~,
\end{equation}
and in $\kappa$-Minkowski, reflecting the deformed symmetries,\footnote{The
symmetries of $\kappa$-Minkowski can be described~\cite{dsr1dsr2} using as
key ingredient a $\kappa$-Poincar\'{e} Hopf algebra. The relevant $\kappa$-Poincar\'{e}
mass-casimir relation is ${\mathcal C}_{\kappa}(k)=m^2$. The mass parameter
$m$, which here appears also in the Lagrangian, is not to be identified
with the rest energy $E(\vec{k}=0)$.
The rest energy $M$ is obtained from $m$ through the
relation~\cite{kpoinap} $m^2 = \kappa^{2} sinh^2(M/(2 \kappa))$.
(Note that however $m$ and $M$ differ only at order $1/\kappa^2$.)}
\begin{equation}
{\mathcal C}_{\kappa}(k) =
\kappa^{2}(e^{k_0/\kappa}+e^{-k_0/\kappa}-2)-\vec{k}^2
e^{-k_0/\kappa}~. \label{casimir}
\end{equation}
It is convenient to introduce the normalized partition function
$\bar{Z}^0\left[J \right] \equiv Z^0\left[J \right] /
Z^0\left[0\right]$, and from (\ref{zeta2nd}) with simple
manipulations one finds that
\begin{equation}
\bar{Z}^0_D\left[J \right]=\exp \left(-\frac{i}{2}\int d^4k ~
\chi_1(k_0) \frac{J(k)J(\dot{-}k)}{{\mathcal C}_{D}(k)-m^2}\right)
~. \label{zeta3}
\end{equation}

To obtain Green functions from (\ref{zeta3}) in the
$\kappa$-Minkowski case one also needs
appropriately generalized definitions of the functional
derivatives:
\begin{equation}
\frac{\delta F\left[f\right]}{\delta f(k)}
=\lim_{\varepsilon\rightarrow 0}\frac{1}{\varepsilon}
\left(F\left[f(p)+\varepsilon\delta^{(4)}_{D}(p\,\dot{+}(\dot{-}k))\right]-F\left[f(p)\right]
\right) ~, \label{functder1}
\end{equation}
\begin{equation}
\frac{\delta F\left[f\right]}{\delta f(\dot{-}k)}
=\lim_{\varepsilon\rightarrow 0}\frac{1}{\varepsilon}
\left(F\left[f(p)+\varepsilon\delta^{(4)}_{D}(p\,\dot{+}k)\right]-F\left[f(p)\right]
\right) ~. \label{functder2}
\end{equation}
Using (\ref{deltaone}), (\ref{functder1}), (\ref{functder2}) and
the property ${\mathcal C}_{D}(\dot{-}p)={\mathcal C}_{D}(p)$,
from
\begin{equation}
G^{(2)}_{0}(p\,,\dot{-}p')=-\left. \frac{\delta^2
\bar{Z}^0\left[J \right]} {\delta J(\dot{-}p)\delta
J(p')}\right|_{J=0} \label{defprop}
\end{equation}
one easily obtains the two-point function at tree-level:
\begin{equation}
G^{(2)}_{0}(p\,,\dot{-}p')=\frac{i}{2} \chi_1(p_0)\chi_1(-p_0)
\frac{\delta^{(4)}(p\dot{+}(\dot{-}p')) +
\delta^{(4)}((\dot{-}p')\dot{+}p)} {{\mathcal C}_{D}(p)-m^2} ~.
\label{prop}
\end{equation}
 From the last equation it is easy to verify that in the canonical case the
two-point function at tree-level is unmodified.\footnote{This could be
guessed already at the level of the generating functional, using the fact
that the antisymmetry of $\theta_{\mu\nu}$ leads to cyclic dependence of
the integrals of product of fields on the order of the fields, and in
particular the product of two fields under integral is undeformed.} On the
$\kappa$-Minkowski side a key role is played by the mass casimir ${\mathcal
C}_{\kappa}$, as one might have expected. Importantly, also on the
$\kappa$-Minkowski side the two $\delta^{(4)}$ in (\ref{prop}) enforce the
same trivial conservation condition; in fact,
$\delta^{(4)}((\dot{-}p')\dot{+}p)
=e^{3p_0/\kappa}\delta^{(4)}(p\dot{+}(\dot{-}p'))
=e^{3p_0/\kappa}\delta^{(4)}(p-p')$. This is a first nontrivial
and reassuring result obtained~\cite{gacmichele} by our approach to field
theory in these noncommutative spacetimes: in $\kappa$-Minkowski spacetime
we find that, in spite of the nonsymmetric and nonlinear coproduct
structure, the usual property that energy-momentum is conserved along the
tree-level two-point function is maintained.

In order to be able to investigate the properties of the two-point function
beyond tree level and in order to establish the form of the tree-level
vertex we must now analyze the $O(\lambda)$ contributions to the
Green functions. For this we must of course reinstate $\lambda \ne 0$, {\it
i.e.} we need to analyze $\bar{Z}^{(1)}_D\left[J \right]$ rather than
$\bar{Z}^0_D\left[J \right]$. It turns out to be useful to rely on some
simple relations between $\bar{Z}^{(1)}_D\left[J \right]$ and
$\bar{Z}^0_D\left[J \right]$, which one can obtain with manipulations
analogous to the ones described above. For the canonical case one finds
\begin{eqnarray}
\bar{Z}^{(1)}_{\theta}\left[J \right] &=&-i\frac{\lambda}{24}
\int \delta^{(4)}\left( \sum_{i=1}^4 k_i \right)
\prod_{j=1}^4\frac{d^4k_j}{2\pi} \exp \left( -\frac{i}{2} \sum_{1
\leq i < j \leq 4} k_{i} {\times}k_{j} \right) {\cdot} \nonumber
\\ &&{\cdot} \frac{\delta}{\delta J(-k_j)}
\bar{Z}^0_{\theta}\left[J \right]\quad,
\end{eqnarray}
where we introduced the notation $p{\times}q=p^{\mu} \theta_{\mu\nu} q^{\nu}$,
while for the $\kappa$-Minkowski case one finds
\begin{equation}
\bar{Z}^{(1)}_{\kappa}\left[J \right] =-i\frac{\lambda}{24}
\int\ \delta^{(4)}\left(\dot{\sum}_{k_1,k_2,k_3,k_4}\right)
\prod_{j=1}^4\frac{d^4k_j}{2\pi} \xi(k_{j,0})\frac{\delta}{\delta
J(\dot{-}k_j)} \bar{Z}^0_{\kappa}\left[J \right] ~,
\label{zeta1stzwei}
\end{equation}
where $\xi(k_{j,0}) \equiv 2
\left(\mu(k_{j,0})+\mu(-k_{j,0})e^{\frac{3k_{j,0}}{\kappa}}\right)^{-1}
$, and we introduced a compact ordered-sum notation:
\begin{equation}
\dot{\sum}_{k_1,k_2,k_3,k_4} \equiv k_1 \dot{+} k_2 \dot{+} k_3 \dot{+}
k_4\quad.
\end{equation}
The $O(\lambda)$ contribution to the two-point function can be written as
\begin{equation}
  G^{(2)}_{\lambda}(p\,,\dot{-}p') \!\!=\!\!
\left(\left.-\frac{\delta^2 \bar{Z}^1_D\left[J \right]}{\delta
J(\dot{-}p)\delta J(p')}\right|_{J=0}\right)_{connected}.
\end{equation}
In both the canonical and the $\kappa$-deformed case one finds that the 24
connected elements of the one loop deformed two-point function split into
two different classes: 16 ``planar" contributions and 8 ``non-planar"
contributions. Planar contributions are associated with the 16
possibilities for attaching the external momenta to consecutive internal
lines ({\it e.g.} to $k_1,k_2$).
The difference between these ``planar" diagrams and the
diagrams, which can be described as ``non-planar", that correspond to the
remaining 8 permutations, in which instead the external lines are attached
to non-consecutive lines, is meaningful in our noncommutative spacetimes.
In fact, on the canonical side the order of the lines coming out of a
vertex is reflected in the structure of the $\theta$-dependent phase
factors, while on the $\kappa$-Minkowski side the coproduct sum
$\dot{\sum}_{k_1,k_2,k_3,k_4}$ is not invariant under $k_1,k_2,k_3,k_4$
permutations. On the canonical side an example of planar contribution is
\begin{eqnarray}
 &&i\frac{\lambda}{24} \int
\prod_{j=1}^4\frac{d^4k_j}{2\pi}
\delta^{(4)}\left( \sum_{i=1}^4 k_i \right)\exp
\left[-\frac{i}{2}\left( p{\times}p'+2p{\times}k_1+2p'{\times}k_1  \right)\right] {\cdot} \nonumber
\\ &&{\cdot}\left. \frac{\delta^2
\bar{Z}^0_{\theta}\left[J \right]}{\delta J(-p)\delta
J(-k_2)}\right |_{J=0} \left. \frac{\delta^2
\bar{Z}^0_{\theta}\left[J \right]}{\delta J(p')\delta J(-k_3)} \right|_{J=0}
\left. \frac{\delta^2 \bar{Z}^0_{\theta}\left[J \right]} {\delta
J(-k_1)\delta J(-k_4)}\right|_{J=0},
\label{eq:spa1}
\end{eqnarray}
and an example of non-planar one is
\begin{eqnarray}
 &&i\frac{\lambda}{24} \int\
\prod_{j=1}^4\frac{d^4k_j}{2\pi}
\delta^{(4)}\left( \sum_{i=1}^4 k_i\right)
\exp \left[-\frac{i}{2}\left( p{\times}p'+2p{\times}k_1  \right)\right]{\cdot}
\nonumber \\ &&{\cdot}\left.\frac{\delta^2
\bar{Z}^0_{\theta}\left[J \right]}{\delta J(-p)\delta
J(-k_2)}\right|_{J=0}\left. \frac{\delta^2
\bar{Z}^0_{\theta}\left[J \right]}{\delta J(p')\delta J(-k_4)}
\right|_{J=0}\left. \frac{\delta^2 \bar{Z}^0_{\theta}\left[J \right]} {\delta
J(-k_1)\delta J(-k_3)}\right |_{J=0}.
\label{eq:spa2}
\end{eqnarray}
 From (\ref{defprop}) and (\ref{prop}) it is possible to see that these
expressions contain the term $\delta^{(4)}(p-p')$, which, in light of the
antisymmetry of $\theta_{\mu\nu}$, allows one to ignore terms
like $p{\times}p'$ and $2p{\times}k_1+2p'{\times}k_1$, and leads tot he conclusion
that planar terms do not
involve any nontrivial $\theta$-dependent phase factors. Nonplanar terms
instead do involve nontrivial phase factors of the type $\exp ({\pm}ip{\times}q)$,
where $p$ and $q$ represent one external and one internal momentum, and the
sign of the exponent depends on the specific momenta involved. Using these
observations, and the result on the tree-level two-point function reported
above, it is easy to combine all the $O(\lambda)$ (tadpole) contributions
(planar and nonplanar) to the full two-point function in the canonical
theory. For the truncated two-point function the result is
\begin{equation}
-i \frac{\lambda}{6}
\int \frac{d^4 k}{(2 \pi)^4} \left( 2+\cos \left( p {\times} k\right)\right)
 \frac{i}{k^2 -m^2}\quad,
\label{canofull}
\end{equation}
where $p$ is the external momentum (the propagation momentum).

In $\kappa$-Minkowski the qualitative (diagrammatic)
description of planar and nonplanar contributions
is completely analogous, but of course the integrals that
are represented by those diagrammatic rules are significantly
different. An example of planar contribution is
\begin{eqnarray}
&&i\frac{\lambda}{24} \int\
\prod_{j=1}^4\frac{d^4k_j}{2\pi}\xi(k_{j,0})
\delta^{(4)}\left(\dot{\sum}_{k_1,k_2,k_3,k_4}\right){\cdot}\nonumber \\
 &&{\cdot}\left.\frac{\delta^2
\bar{Z}^0_{\kappa}\left[J \right]}{\delta J(\dot{-}p)\delta
J(\dot{-}k_2)}\right |_{J=0}\left. \frac{\delta^2
\bar{Z}^0_{\kappa}\left[J \right]}{\delta J(p')\delta J(\dot{-}k_3)} \right
|_{J=0}\left. \frac{\delta^2 \bar{Z}^0_{\kappa}\left[J \right]} {\delta
J(\dot{-}k_1)\delta J(\dot{-}k_4)}\right |_{J=0},
\label{eq:spa3}
\end{eqnarray}
while an example of non-planar contribution is
\begin{eqnarray}
&&i\frac{\lambda}{24} \int\
\prod_{j=1}^4\frac{d^4k_j}{2\pi}\xi(k_{j,0})
\delta^{(4)}\left(\dot{\sum}_{k_1,k_2,k_3,k_4}\right){\cdot} \nonumber \\
 &&{\cdot}\left. \frac{\delta^2
\bar{Z}^0_{\kappa}\left[J \right]}{\delta J(\dot{-}p)\delta
J(\dot{-}k_2)}\right |_{J=0}\left. \frac{\delta^2
\bar{Z}^0_{\kappa}\left[J \right]}{\delta J(p')\delta J(\dot{-}k_4)} \right
|_{J=0}\left. \frac{\delta^2 \bar{Z}^0_{\kappa}\left[J \right]} {\delta
J(\dot{-}k_1)\delta J(\dot{-}k_3)}\right |_{J=0}.
\label{eq:spa4}
\end{eqnarray}
Before looking at how all the planar and all the nonplanar contributions
combine in this case to give the $O(\lambda)$ expression of the full two
point function in $\kappa$-Minkowski, let us pause on these rather implicit
formulas for the $\kappa$-Minkowski tadpole contributions and compare them
to the corresponding formulas for the case of canonical spacetimes. One
should notice that:
\begin{description}
\item[(i)] The fact that the tree-level two
point functions are different also leads (since the tree-level two-point
function appears in the integrand) to significant differences between the
$\kappa$-Minkowski case and the canonical case for the computation of the
tadpole contributions.
\item[(ii)] It is also important that in the
canonical case the integration measure is trivial (just as in commutative
spacetimes), whereas in the $\kappa$-Minkowski case the measure is
nontrivial (and the fact that, at present, there is no consensus on the
choice of this measure imposes severe limitations on what can be reliably
computed in $\kappa$-Minkowski field theories).
\item[(iii)] The spacetime noncommutativity is reflected in
energy-momentum space through nontrivial phase factors, on the canonical
side, and through nontrivial conservation laws (delta functions) on the
$\kappa$-Minkowski side.
\item[(iv)] Planar diagrams do not carry any trace of the nontrivial
structures mentioned in the previous point. We have already discussed the
fact that in the canonical case there is no nontrivial phase factor for the
planar contributions, and from (\ref{eq:spa3}) one can easily
verify~\cite{gacmichele} that the deformed delta functions of planar
tadpole diagrams in $\kappa$-Minkowski combine to ultimately impose
trivial/ordinary energy-momentum conservation. [For example, one finds that
(\ref{eq:spa3}) is proportional to $
\delta^{(4)}(\dot{-}k_4\dot{+}(\dot{-}p)\dot{+}p'\dot{+}k_4)\sim
\delta^{(4)}(p-p')$.]
\item[(v)] Nonplanar diagrams encode the most important new features induced
by spacetime noncommutativity.\footnote{This should not surprise the
readers: the difference between planar and nonplanar diagrams is meaningful
because the order of lines attached around a vertex is itself meaningful,
and this has its root in the fact that fields do not commute ({i.e.} it
originates directly from noncommutativity).} We have already discussed the
fact that in the canonical case nonplanar contributions do involve nontrivial
phase factors, and from (\ref{eq:spa4}) one can easily
verify~\cite{gacmichele} that the deformed delta functions of nonplanar
tadpole diagrams in $\kappa$-Minkowski combine to ultimately impose
deviations from ordinary energy-momentum conservation. For example, one
finds that (\ref{eq:spa4}) is proportional to
$\delta^{(4)}(\dot{-}k_3\dot{+}(\dot{-}p)\dot{+}k_3\dot{+}p')\sim
 \delta(p_0-p'_0)\,\,\delta^{(3)}(e^{p_0/\kappa}\vec{k_3}-
\vec{p}+\vec{k_3}+ e^{k_{3,0}/\kappa}\vec{p'})$.
\end{description}
These five points {\bf (i)-(v)} summarize the main differences in the
tadpole structure on the two sides, which are representative of the
differences that one encounters in the general Feynman-diagram analysis of
these theories. It is worth making a few remarks specifically on the
deviations from ordinary energy-momentum conservation (in propagation!!)
that emerge from the nonplanar tadpole contributions in $\kappa$-Minkowski.
In $\kappa$-Minkowski energy conservation is still ordinary also for
non-planar diagrams, but momentum conservation is modified and it is
modified in a way that cannot even be described as a modified conservation
law: the terms involving the loop/integration momenta do not fully cancel
each other out in the argument of the left-over (conservation-imposing)
delta function. It is easy to verify that these non-planar contributions,
while not implementing exactly the ordinary energy-momentum conservation,
are still mainly centered around ordinary energy-momentum conservation
(assuming reasonably good behaviour at infinity of the expression under the
integral). There is therefore some deviation from ordinary conservation of
energy-momentum, a sort of fuzzy conservation of momentum, but it is
plausible that the full theory (whose construction will also require the
integration measure that we are here treating as an unknown) would only
predict a very small ($1/\kappa$-suppressed) deviation from ordinary
conservation of energy-momentum, possibly consistent with observational
limits. This issue is here postponed to future studies.

Having noted these main qualitative features of tadpole contributions on
the two sides, before moving on to the analysis of interaction vertices,
let us note here some formulas that describe the full (truncated) tadpole
in $\kappa$-Minkowski. This would be the $\kappa$-Minkowski result that
corresponds to the result (\ref{canofull}) for the canonical spacetimes.
While in the canonical case a large number of simplifications could be
exploited to obtain a compact formula, in the $\kappa$-Minkowski case one
is stuck with a very long formula which we here split for convenience in
two pieces: the sum of the contributions to the
truncated tadpole that come from planar diagrams
\begin{eqnarray}
&&i\frac{\lambda}{24}\mu(-p_0)\mu(-p_0)\xi^2(p_0)
\delta^{(4)}(p'-p){\cdot}\nonumber\\
&&{\cdot}2\left(1+e^{\frac{p_0}{\kappa}}\right)
\left(\int\frac{d^4k_4}{(2\pi)^4}\xi^2(k_{0,4})\mu^2(-k_{0,4})\mu(k_{0,4})
e^{\frac{k_{0,4}}{\kappa}}
\frac{e^{-\frac{k_{0,4}}{\kappa}}+1}{{\mathcal
C}_{\kappa}(k_4)-m^2}+\right.\nonumber\\ &&
\left.\sum_{j=1}^3\int\frac{d^4k_j}{(2\pi)^4}
\xi^2(k_{0,j})\mu^2(-k_{0,j})\mu(k_{0,j})
\frac{e^{-\frac{k_{0,j}}{\kappa}}+1}{{\mathcal
C}_{\kappa}(k_j)-m^2}\right)\, , \label{ktadp}
\end{eqnarray}
and the sum of the contributions that come from nonplanar diagrams
\begin{eqnarray}
&&i\frac{\lambda}{24}\mu(-p_0)\mu(-p_0)\xi^2(p_0){\cdot}\nonumber\\
&&{\cdot}2\left(\int\frac{d^4k_3}{(2\pi)^4}\xi^2(k_{0,3})\mu^2(-k_{0,3})\mu(k_{0,3})
\delta^{(4)}(\dot{-}k_3\dot{+}(\dot{-}p)\dot{+}k_3\dot{+}p')
\frac{e^{-\frac{k_{0,3}}{\kappa}}+1}{{\mathcal
C}_{\kappa}(k_3)-m^2}+\right.\nonumber\\
&&+\int\frac{d^4k_3}{(2\pi)^4}\xi^2(k_{0,3})\mu^2(-k_{0,3})\mu(k_{0,3})
\delta^{(4)}(\dot{-}k_3\dot{+}p'\dot{+}k_3\dot{+}(\dot{-}p))
\frac{e^{-\frac{k_{0,3}}{\kappa}}+1}{{\mathcal
C}_{\kappa}(k_3)-m^2}+\nonumber\\
&&+\int\frac{d^4k_2}{(2\pi)^4}\xi^2(k_{0,2})\mu^2(-k_{0,2})\mu(k_{0,2})
\delta^{(4)}(p'\dot{+}k_2\dot{+}(\dot{-}p)\dot{+}(\dot{-}k_2))
\frac{e^{-\frac{k_{0,2}}{\kappa}}+1}{{\mathcal
C}_{\kappa}(k_2)-m^2}+\nonumber\\
&&+\left.\int\frac{d^4k_2}{(2\pi)^4}\xi^2(k_{0,2})\mu^2(-k_{0,2})\mu(k_{0,2})
\delta^{(4)}(\dot{-}p\dot{+}k_2\dot{+}p'\dot{+}(\dot{-}k_2))
\frac{e^{-\frac{k_{0,2}}{\kappa}}+1}{{\mathcal
C}_{\kappa}(k_2)-m^2}\right)\, .\nonumber\\ \label{ktadnp}
\end{eqnarray}

Our next, and final, task is the study of the tree-level vertex.
The $O(\lambda)$ contribution to the four-point Green function can
be expressed in terms of $\bar{Z}^{(1)}$ through
\begin{equation}
G^{(4)}_{\lambda}(p_1,p_2,\dot{-}p_3,\dot{-}p_4)=
\left. \frac{\delta^4 \bar{Z}^{(1)}_D\left[J \right]}{\delta J(\dot{-}p_1)\delta
 J(\dot{-}p_2)\delta J(p_3)\delta J(p_4)}\right |_{J=0} ~,
\label{fourpf}
\end{equation}
and the tree-level (untruncated)
vertex is the sum of connected graphs contributing to
$G^{(4)}_{\lambda}(p_1,p_2,\dot{-}p_3,\dot{-}p_4)$. The contributions in
the two cases we are analyzing are very different. For example one of the
contributions\footnote{The contribution
reported in Eq.~(\ref{fourpf2c})
is actually one of the terms in which the phase factor
containing $\theta_{\mu\nu}$ takes its simplest form. In other
cases the six terms of type $p_i {\times}p_j$ in the exponent
have different signs.} for the canonical case turns out to have the form
\begin{eqnarray}
&&-\frac{i\lambda}{24} \int \left( \prod_{l=1}^4
\frac{d^4k_l}{2\pi}\right) \delta^{(4)} \left( \sum_{i=1}^4
k_i\right)\exp \left(-\frac{i}{2} \sum_{1 \leq i < j \leq 4} p_i
{\times}p_j  \right)\left. \frac{\delta^2
\bar{Z}^0_{\theta}\left[J \right]}{\delta J(-p_{1})\delta
J(-k_1)}\right |_{J=0}{\cdot} \nonumber \\ && {\cdot}\left.
\frac{\delta^2 \bar{Z}^0_{\theta}\left[J \right]}{\delta
J(-p_{2})\delta J(-k_2)}\right|_{J=0}\left.\frac{\delta^2
\bar{Z}^0_{\theta}\left[J \right]}{\delta J(p_3)\delta J(-k_3)}
\right |_{J=0}\left. \frac{\delta^2
\bar{Z}^0_{\theta}\left[J \right]}{\delta J(p_4)\delta J(-k_4)}
\right|_{J=0}\quad, \label{fourpf2c}
\end{eqnarray}
while the same term in the $\kappa$-Minkowski case
takes the form
\begin{eqnarray}
&&-\frac{i\lambda}{24} \int \left( \prod_{l=1}^4
\frac{d^4k_l}{2\pi}\xi(k_{l,0})\right)
\delta^{(4)}\left(\dot{\sum}_{k_1,k_2,k_3,k_4}\right)
\left.\frac{\delta^2 \bar{Z}^0_{\kappa}\left[J \right]}{\delta
J(\dot{-}p_{1})\delta J(\dot{-}k_1)}\right |_{J=0}
{\cdot}\nonumber
\\ && {\cdot}
\left. \frac{\delta^2
\bar{Z}^0_{\kappa}\left[J \right]}{\delta J(\dot{-}p_{2})\delta
J(\dot{-}k_2)}\right |_{J=0}
\left.\frac{\delta^2 \bar{Z}^0_{\kappa}\left[J \right]}{\delta J(p_3)\delta
J(\dot{-}k_3)} \right |_{J=0} \left. \frac{\delta^2
\bar{Z}^0_{\kappa}\left[J \right]}{\delta J(p_4)\delta J(\dot{-}k_4)} \right
|_{J=0} ~.
\label{fourpf2d}
\end{eqnarray}
We are primarily interested in establishing what are the conservation rules
implemented at the vertex. Clearly for the case of canonical noncommutative
spacetime one finds the same ordinary energy-momentum conservation rule
that applies in the familiar commutative Minkowski spacetime. It is again
in the $\kappa$-Minkowski noncommutative spacetime that the most dramatic
new features emerge. It is important to observe that (\ref{fourpf2d}) is
proportional to $\delta^{(4)}(\dot{-}p_1
\dot{+}(\dot{-}p_2)\dot{+}p_3\dot{+}p_4)$ which corresponds to ordinary
energy conservation, $-p_{1,0}-p_{2,0}+p_{3,0}+p_{4,0}=0$, but enforces a
non-trivial and non-symmetric rule of conservation of 3-momenta:
$-e^{\frac{-p_{1,0}}{\kappa}}\vec{p}_1-e^{-\frac{1}{\kappa}
(p_{1,0}+p_{2,0})}\vec{p}_2+e^{-\frac{1}{\kappa}(p_{1,0}+p_{2,0})}
\vec{p}_3 +e^{\frac{1}{\kappa}(-p_{1,0}-p_{2,0}+p_{3,0})} \vec{p}_4=0$.
There are 23 other contributions, analogous to (\ref{fourpf2d}),
to the $\kappa$-Minkowski vertex, and all are proportional
to a delta function of the
type $\delta^{(4)}(\dot{-}p_1 \dot{+}(\dot{-}p_2)\dot{+}p_3\dot{+}p_4)$
but with different ordering of
the momenta $\dot{-}p_1,\dot{-}p_2,p_3,p_4$.
By using the properties of the coproduct/antipode ($\dot{+}/\dot{-}$)
it is easy to see that different ordering possibilities
within the argument of the delta function lead to inequivalent
conservation rules.
This is completely different from the behaviour of the
$\kappa \rightarrow \infty$ limit (the limit in which $\kappa$-Minkowski
reduces to the ordinary commutative Minkowski spacetime),  in
which all 24 contributions lead to the same
conservation rule $\delta^{(4)}(-p_1-p_2+p_3+p_4)$.

Before making additional remarks
it is useful to note here
the formulas for the full truncated vertex
(putting together the 24 contributions on each side),
which in the canonical case takes the form
\begin{eqnarray}
\frac{-i\lambda}{3}&&\left( \cos \left( \frac{k_1{\times}k_4-k_2{\times}k_3}{2} \right)
+\cos \left( \frac{k_1{\times}k_3+k_2{\times}k_4}{2} \right)+\right.\nonumber\\
&&+\left.\cos\left( \frac{k_1{\times}k_2-k_3{\times}k_4}{2} \right)\right) \delta^{(4)}
\left(\sum_{i=1}^4 k_i\right)\quad,
\end{eqnarray}
while in $\kappa$-Minkowski it takes the form
\begin{eqnarray}
&&i\frac{\lambda}{24}\left(\frac{i}{2}\right)^4
\mu(-p_{0,1})\mu(-p_{0,2})\mu(p_{0,3})
\mu(p_{0,4})\left(\prod_{j=1}^4\xi(-p_{0,j})\right){\cdot}\nonumber\\
&&{\cdot}\left(\delta^{(4)}(\dot{-}p_1\dot{+}(\dot{-}p_2)\dot{+}p_3\dot{+}p_4)+
{\mathcal{P}}_{\dot{-}p_1,\,\dot{-}p_2,\,\dot{+}p_3,\,\dot{+}p_4}\right)\,
, \label{kfourpf}
\end{eqnarray}
where $\mathcal{P}$ encodes the 24 permutations described above. It is
important to notice that, in spite of the lack of symmetry under exchange
of momenta that one encounters at intermediate stages of the analysis, the
overall structure of the interaction vertices is fully symmetric under
exchanges of entering momenta. On the $\kappa$-Minkowski side we are
however confronted with a revision of the concept of energy-momentum
conservation for scattering processes: since our vertex is not
characterized by an overall $\delta$-function (but instead it is split up
into different pieces characterized by different $\delta$-function
factors), in a given
scattering process, with incoming particles characterized by four-momenta
$p_1$ and $p_2$, it becomes impossible to predict the sum of the 3-momenta
of the outgoing particles. The theory only predicts that one of our 24
energy-momentum-conservation rules must be satisfied and assigns (equal)
probabilities to each of these 24 channels.\footnote{These properties of
vertices in $\kappa$-Minkowski spacetime represent a rather significant
departure from conventional physics and therefore provide a key tool for
testing whether Nature makes use of $\kappa$-Minkowski. Making the
reasonable assumption~\cite{dsr1dsr2} that $\kappa$ should be of the order
of the Planck scale one easily checks that this prediction for new
(non-)conservation rules at the vertex is consistent with all available
low-energy data. (In the limit $p_0/\kappa \ll 1$ the 24 different
conservation rules that characterize our $\kappa$-deformed vertex collapse
into a single, and ordinary, conservation rule.)}

We stop here with our analysis of these field theories. Of
course, many more issues deserve being discussed: the
contributions to the two-point function that, unlike the tadpole,
involve a genuine flow of external momenta in loops, the issue of
unitarity of the theories, and many other issues. For some of
these issues there is not yet a satisfactory analysis in the
published literature, and particularly on the $\kappa$-Minkowski
side some of these issues present us with overwhelming technical
and conceptual challenges.~\cite{gacmichele} There is strong
motivation for future studies attempting to address these issues.
In fact, these theories in noncommutative spacetimes predict
(unlike most ``quantum pictures" of spacetime) new physical
effects whose magnitude could be large enough for detection (or
rejection) in forthcoming experiments. We close by emphasizing
some aspects of this phenomenological programme.

It should not go unnoticed that this phenomenology, while certainly
interesting, presents us with severe challenges at both the conceptual and
quantitative level. Field theories in canonical noncommutative spacetimes
host a highly nontrivial infrared structure. Some evidence of this can be
seen by calculating the integral in Eq.~(\ref{canofull}) and realizing that
the result has singular behaviour at small external momentum. More careful
discussions of this feature (and of its origin in the so-called ``IR/UV
mixing") can be found, {\it e.g.}, in
Refs.~~\cite{sussIRUV,seibergIRUV,gacluisa}. The nontrivial infrared
structure poses a severe challenge~\cite{gacluisa,gacgianluca} to our
conventional way to do phenomenology (but unfortunately this challenge has
been largely underestimated in the large majority of publications on this
subject).

These infrared problems of the phenomenology
on the canonical side do not appear to
be present on the $\kappa$-Minkowski side. However, there one is confronted
with the fact that the natural estimate of $\kappa$ as the Planck
scale~\cite{dsr1dsr2} (or somewhere in that neighborhood) leads to
the prediction of very small effects. It is natural to identify $\kappa$
with a fundamental/meaningful scale since $\kappa$ is an invariant of the
appropriate formulation~\cite{dsr1dsr2} of Lorentz transformations
that applies in $\kappa$-Minkowski (different inertial observers attribute
the same numerical value to $\kappa$). For the entries in the $\theta$
matrix on the canonical side there is instead no natural estimate, since
their values are observer-dependent (the proper concept of Lorentz
transformations in canonical noncommutative spacetime is just the ordinary
one and under those Lorentz transformations $\theta$
behaves like a tensor, {\it i.e.} its components take different numerical
values for different inertial observers).

Perhaps the most exciting phenomenological developments in this area would
come if (at least in these elementary quantum spacetimes) one could give a
satisfactory description~\cite{gacGWI} of spacetime fuzziness. Experimental
sensitivity to certain possible manifestations of fuzziness is improving
very quickly. However, our present understanding of fuzziness in these
spacetimes is still very limited.~\cite{gacplbkpoin}

At least for the short-term future a more promising opportunity for
experimental tests comes from the emergence of deformed dispersion
relations in these noncommutative spacetimes. We have shown above that the
dispersion relation is modified already at tree level in
$\kappa$-Minkowski. On the canonical side the tree-level dispersion
relation is undeformed, but loop effects introduce severe deformation.
[Again one can get a first glimpse at this feature by calculating the
integral in Eq.~(\ref{canofull}).] Experimental sensitivities to ``in vacuo
dispersion",~\cite{aemn,grbgac} a characteristic signature of deformed
dispersion relations, have improved dramatically in recent times. In
particular, within a few years the $\kappa$-Minkowski dispersion relation
will be tested~\cite{grbgac,billetal,glast} with sensitivity all the way up
to $\kappa$ of order the Planck scale.

Deformed dispersion relations are also being analyzed in relation with the
emergence of deformed threshold conditions for particle-production
processes,~\cite{kifu,aus,gactp,jacobson} another prediction which could be
tested with remarkable sensitivity.~\cite{crdata,mkdata}


\end{document}